\documentstyle[amsfonts,epsf,12pt]{article}
\renewcommand{\d}{{\rm d}}
\newcommand{\e}{{\rm e}}
\begin{document}
\title{Quantum and classical stochastic dynamics:\\
Exactly solvable models by supersymmetric methods\thanks{Based on an invited
talk presented at the \emph{III. International Workshop on Classical and
  Quantum Integrable Systems}, Yerevan, Armenia, June 29 - July 04, 1998.}}
\author{Georg Junker\\[5mm] 
Institut f\"ur Theoretische Physik, Universit\"at Erlangen-N\"urnberg,\\ 
Staudtstr.\ 7, D-91058 Erlangen, Germany\\
\emph{E-mail:} junker@theorie1.physik.uni-erlangen.de}
\date{}
\maketitle

%%%%%%%%%%%%%%%%%%%%%%%%%%%%%%%%%%%%%%%%%%%%%%%%%%%%%%%%%%%%%%%%%%%%%%%%%%%%%%
\begin{abstract}
A supersymmetric method for the construction of so-called conditionally
exactly solvable quantum systems is reviewed and extended to classical
stochastic dynamical systems characterized by a Fokker-Planck equation with
drift. A class of drift-potentials on the real line as well as on
the half line is constructed for which the associated Fokker-Planck equation
can be solved exactly. Explicit drift potentials, which describe mono-, bi-,
meta-or unstable systems, are constructed and their decay rates and modes are
given in closed form.
\end{abstract}
\mbox{}

%%%%%%%%%%%%%%%%%%%%%%%%%%%%%%%%%%%%%%%%%%%%%%%%%%%%%%%%%%%%%%%%%%%%%%%%%%%%%%
\noindent{\large\bf 1\quad Introduction}\\[2mm]
Many physical phenomena of nature are characterized by some basic differential
equations. For example, quantum-mechanical phenomena are described by
Schr\"o\-dinger's equation, which dictates the dynamics of some quantum system
represented by a Hamilton operator. One is therefore primarily interested in
finding all eigenvalues and eigenstates of such Hamiltonians. As a
consequence, finding a large class of, in this sense, analytically exactly
solvable quantum systems is an important goal and this search has already been
initiated by Schr\"odinger using the so-called factorization method
\cite{Schr,IH51}. This factorization method is basically equivalent to
Darboux's method \cite{Da82} already invented 1882 and applied to
Sturm-Liouville problems. Recently, these methods have attracted new attention
\cite{M84}, in particular in connection with supersymmetric (SUSY) quantum
mechanics \cite{Ju96}. For a recent approach and further references see, for
example, \cite{RJ97,RJ98a,RJ98b}. 

Somehow unnoticed by those working on the quantum mechanical problem
the same basic ideas have been utilized in constructing exactly solvable
classical stochastic systems described by a Fokker-Planck equation
\cite{HZ82,HR85,Ja88}. In fact, the Fokker-Planck equation can be 
put into the form of an imaginary-time Schr\"odinger equation exhibiting
SUSY. Therefore, exactly solvable SUSY quantum Hamiltonians also provide
exactly solvable Fokker-Planck-type systems. Whereas until now these methods
have been applied to mono- and bistable systems \cite{HZ82,HR85,Ja88},
they correspond to a situation with unbroken SUSY, we present here also
examples with broken SUSY giving rise to meta- or unstable drift potentials. 

In the next two sections we briefly review the basic concepts of SUSY
quantum mechanics and its close connection with the Fokker-Planck
equation \cite{Ju96,vK92}. Section 4 reviews our recent approach
\cite{RJ97,RJ98a,RJ98b} towards a construction of the most general class of
SUSY partners of a given quantum mechanical Hamiltonian. This approach is
applied to the Fokker-Planck equation and allows to construct new drift
potentials with known decay rates and modes. Two examples are discussed in some
detail. The first one is related to the linear harmonic oscillator having
unbroken SUSY and leads to bistable double-well or monostable single-well
drift potentials already found by Hongler and Zheng \cite{HZ82}. The
second example is the class related to the radial harmonic oscillator
with broken SUSY. Here the drift potential is either metastable or
unstable.\\[2mm]

%%%%%%%%%%%%%%%%%%%%%%%%%%%%%%%%%%%%%%%%%%%%%%%%%%%%%%%%%%%%%%%%%%%%%%%%%%%%%%
\noindent{\large\bf 2\quad Supersymmetric quantum mechanics}\\[2mm]
To begin with, let us briefly review the basics of SUSY quantum
mechanics or, to be more precise, Witten's model of SUSY quantum mechanics
\cite{Ju96}. This model consists of a pair of standard Schr\"odinger
Hamiltonians (units are such that $\hbar=m=1$)
\begin{equation}
H_\pm=-\frac{1}{2}\,\frac{\d^2}{\d x^2}+V_\pm(x)\;,\qquad
V_\pm(x)= \frac{1}{2}\Bigl(W^2(x)\pm W'(x)\Bigr)\;.
\label{2.1}
\end{equation}
Evidently, this pair is completely characterized by the so-called SUSY
potential $W$, which is assumed to be real-valued and an at least once
differentiable 
function inside the configuration space. This configuration space
will be either the real line ${\mathbb R}$ or the positive half line
${\mathbb R}^+$. Thus the above Hamiltonians act on a suitable linear space of
square-integrable functions over ${\mathbb R}$ and ${\mathbb R}^+$,
respectively. With the help of the supercharge operators
\begin{equation}
A=\frac{1}{\sqrt{2}}\left(\frac{\d}{\d x}+W(x)\right)\;,\quad
A^\dagger=\frac{1}{\sqrt{2}}\left(-\frac{\d}{\d x}+W(x)\right)
\label{2.3}
\end{equation}
the above pair of SUSY-partner Hamiltonians factorizes, $H_+=AA^\dagger\geq
0$, $H_-=A^\dagger A\geq 0$, and obviously obeys the intertwining relations 
$AH_-=H_+A$ and $H_-A^\dagger = A^\dagger H_+$. As a consequence
$H_+$ and $H_-$ are essentially isospectral. In other words, their strictly
positive energy eigenvalues coincide and the corresponding eigenstates are
related via the supercharges. However, there may exist an additional
vanishing eigenvalue for one of these Hamiltonians. In this case SUSY is said
to be unbroken and by standard convention \cite{Ju96} this additional
eigenvalue is assumed to belong to $H_-$. If such a state does not exist SUSY
is said to be broken. To summarize, for unbroken SUSY we have 
\begin{equation}
 \begin{array}{l}
  \displaystyle
  E_0^-=0\;,\qquad  E_{n+1}^-=E_n^+>0\;, \qquad
  \psi_0^-(x)=\psi_0^-(0)\exp\left\{-\int_0^x \d z\,W(z)\right\}\;,\\[2mm]
  \psi_{n+1}^-=(E_n^+)^{-1/2}A^\dagger\psi_n^+\;,\qquad
\psi_{n}^+=(E_{n+1}^-)^{-1/2}A\psi_{n+1}^-\;,
 \end{array}
\label{trafounbroken}
\end{equation}
whereas for broken SUSY these relations read
\begin{equation}
E_{n}^-=E_n^+>0\;,\qquad
\psi_{n}^-=(E_n^+)^{-1/2}A^\dagger\psi_n^+\;,\qquad
\psi_{n}^+=(E_{n}^-)^{-1/2}A\psi_{n}^-\;.
\label{trafobroken}
\end{equation}
Here we have denoted the eigenfunctions and eigenvalues of
$H_\pm$ by $\psi^\pm_n$ and $E^\pm_n$, respectively. That is,
\begin{equation}
H_\pm\psi^\pm_n=E^\pm_n\psi^\pm_n\;,\qquad n=0,1,2,\ldots,
\label{2.5}
\end{equation}
and we also note that throughout this paper we will consider, without loss of
generality, quantum systems with a purely discrete spectrum.

We conclude this section by noting that for broken as well as unbroken SUSY
one can obtain the complete spectral information of one Hamiltonian, say
$H_-$, if the eigenvalues and eigenstates of the corresponding partner, here
$H_+$, and the SUSY potential $W$ are known.\\[2mm]

%%%%%%%%%%%%%%%%%%%%%%%%%%%%%%%%%%%%%%%%%%%%%%%%%%%%%%%%%%%%%%%%%%%%%%%%%%%%%%
\noindent{\large\bf 3 \quad Classical stochastic dynamics}\\[2mm]
As mentioned in the Introduction we are also interested in classical
systems with a stochastic dynamics governed by the Fokker-Planck equation
\begin{equation}
  \frac{\partial}{\partial t}\,m_t(x,x_0)=
  \frac{1}{2}\,\frac{\partial^2}{\partial x^2}\, m_t(x,x_0)
  +\frac{\partial}{\partial x}\Bigl(U'(x) m_t(x,x_0)\Bigr)\;.
\label{FPeq}
\end{equation}
Here $m_t(x,x_0)$ denotes the transition-probability density of a macroscopic
degree of freedom to be found at time $t$ at position $x$ if it initially has
been at $x_0$, i.e.\ $m_0(x,x_0)=\delta(x-x_0)$. This degree of freedom is
subjected to an external 
force characterized by the real-valued drift potential $U$ and to a stochastic
random force (white noise) resulting in the diffusive term on the right-hand
side of (\ref{FPeq}) with diffusion constant set equal to $1/2$.
Making the ansatz \cite{vK92}
\begin{equation}
m_t(x,x_0)=\e^{-U(x)} K_t(x,x_0)
\end{equation}
leads to
\begin{equation}
  -\frac{\partial}{\partial t}\,K_t(x,x_0)=
  \left[-\frac{1}{2}\,\frac{\partial^2}{\partial x^2}+\frac{1}{2}\,U'^{\,2}(x)-
   \frac{1}{2}\,U''(x)\right] K_t(x,x_0)\;,
\end{equation}
which may be interpreted as an imaginary-time Schr\"odinger equation for the
SUSY Hamiltonian $H_-$ with a SUSY potential given by the first derivative of
the drift potential, $W=U'$. Hence, the desired transition-probability density
is given via the Euclidean propagator for $H_-$:
\begin{equation}
m_t(x,x_0)=\exp\{U(x_0)-U(x)\} \langle x|\exp\{-tH_-\}|x_0\rangle\;.
\end{equation}
It is obvious that the decay modes and decay rates of this classical
stochastic dynamical system are related to the eigenfunctions and eigenvalues
of $H_-$. To be more explicit, let us consider the cases of unbroken and broken
SUSY separately.

For unbroken SUSY the ground-state energy of $H_-$ vanishes and as a
consequence there exists a stationary, i.e.\ time-independent, probability
distribution. Noting that $\psi_0^-(x)=\psi_0^-(x_0)\exp\{U(x_0)-U(x)\}$ the
transition-probability density can be put into the form
\begin{equation}
m_t(x,x_0)= \left[\psi_0^-(x)\right]^2
+ \exp\{U(x_0)-U(x)\}\sum_{n=1}^\infty \e^{-tE_n^-} \psi_n^-(x)\psi_n^-(x_0)\;,
\end{equation}
which clearly shows that the strictly positive eigenvalues of $H_-$ and the
associated eigenfunctions characterize the decay rates and decay modes of the
system. In addition the ground-state wavefunction represents the stationary
distribution. That is, a stable system is characterized by an unbroken SUSY.

In the case of broken SUSY, by definition, $H_-$ does have strictly positive
eigenvalues only and cannot lead to a stationary distribution. In other 
words, broken SUSY is related to unstable systems. Here the transition
probability density has the form
\begin{equation}
m_t(x,x_0) = \exp\{U(x_0)-U(x)\}
\sum_{n=0}^\infty \e^{-tE_n^-} \psi_n^-(x)\psi_n^-(x_0)\;.
\end{equation}

Finally, we mention that in both cases we may invert the drift potential,
$U\to-U$, which amounts in replacing $H_-$ by $H_+$. Hence, due to the
presence of SUSY the inverted drift potential has the same decay rates as the
original one. Only in the case of unbroken SUSY a stable system becomes
unstable upon inversion. This is expressed in the fact that the vanishing
eigenvalue $E_0^-=0$ is missing in the spectrum of $H_+$. The decay modes of
the original and the inverted drift potential are also clearly related by the
SUSY transformations (\ref{trafounbroken}) and (\ref{trafobroken}),
respectively. \\[2mm]

%%%%%%%%%%%%%%%%%%%%%%%%%%%%%%%%%%%%%%%%%%%%%%%%%%%%%%%%%%%%%%%%%%%%%%%%%%%%%%
\noindent{\large\bf 4 \quad Designing exactly solvable models}\\[2mm]
From the discussion in the previous two sections it is clear that once the
spectral data, that is, the eigenvalues and eigenfunctions, of $H_+$ are known
we also know these data for the corresponding superpartner $H_-$ and in turn
the decay rates and modes for the Fokker-Planck equation with drift potential 
$U(x)=\int_{x_0}^x \d z\,W(z)$. In order to construct new exactly solvable
systems we will search for the most general class of superpartners associated
with a given exactly solvable quantum Hamiltonian $H_+$. Such a construction
method has recently been given \cite{RJ97,RJ98a,RJ98b} and we briefly review
it here.

In order to find the most general class of SUSY partners of a given
SUSY Hamiltonian we make the following ansatz
\begin{equation}
W(x)=\Phi(x)+\frac{u'(x)}{u(x)}\;,
\label{W}
\end{equation}
where $\Phi$ is a known SUSY potential belonging to some shape-invariant
(i.e.\ exactly solvable) potential. See, for example, Table 5.1.\ in ref.\
\cite{Ju96}. If we now assume that $u$ is a solution of
\begin{equation}
u''(x)+2\,\Phi(x)\,u'(x)-b\,u(x)=0\;,\qquad b\in{\mathbb R}\;,
\label{ODEforu}
\end{equation}
we find that 
\begin{equation}
  V_+(x)=\frac{1}{2}\,\Phi^2(x)+\frac{1}{2}\,\Phi'(x)+\frac{b}{2}\;.
\end{equation}
By construction $V_+$ is up to the additive constant $b/2$ shape-invariant and,
therefore, the eigenvalues and eigenfunctions of $H_+$ are exactly known.
The corresponding partner potential can be put into the form 
\begin{equation}
  V_-(x)=\frac{1}{2}\,\Phi^2(x)-\frac{1}{2}\,\Phi'(x)+\frac{u'(x)}{u(x)}\left(
2\,\Phi(x)+\frac{u'(x)}{u(x)}\right) -\frac{b}{2}\;,
\end{equation}
which, in general, is not shape-invariant and thus a \emph{new} exactly
solvable quantum mechanical potential. Similarly we may construct also a
\emph{new} drift
potential 
\begin{equation}
  U(x)=\int_{x_0}^x\d z\,\Phi(z) + \log u(x)
\end{equation}
for which the associated decay rates and modes are known exactly.

Let us note here that we cannot take an arbitrary solution of (\ref{ODEforu})
for the construction of new quantum-mechanical and drift potentials. In order
to circumvent domain questions of the operators involved we will take only
those solutions of (\ref{ODEforu}) into account which are strictly
positive. Hence, no singularities inside the configuration space will appear
in $V_-$ and $U$. For this reason the new Schr\"odinger potentials $V_-$
obtained in this way have been called conditionally exactly solvable
\cite{RJ97}. In the following 
two subsections we will present two examples with unbroken and broken
SUSY, respectively. For further details and examples see \cite{RJ98b}.\\[2mm]

%%%%%%%%%%%%%%%%%%%%%%%%%%%%%%%%%%%%%%%%%%%%%%%%%%%%%%%%%%%%%%%%%%%%%%%%%%%%%%
\noindent{\bf 4.1 \quad A system with unbroken SUSY}\\[2mm]
The first example we are going to present is characterized by a linear SUSY
potential, $\Phi(x)=x$, with configuration space given by the real line,
$x\in{\mathbb R}$. This SUSY potential gives rise to the harmonic-oscillator
potential 
\begin{equation}
V_+(x)=\textstyle\frac{1}{2}\left(x^2+b+1\right)
\end{equation}
and the corresponding eigenvalues and eigenfunctions of $H_+$ are
\begin{equation}
E_n^+=n+b/2 +1\;,\qquad
\psi_n^+(x)=\left[\sqrt{\pi}\,2^nn!\right]^{-1/2}H_n(x)\exp\{-x^2/2\}\;.
\end{equation}
Here $H_n$ denotes a Hermite polynomial of order $n$. In this case the general
solution of (\ref{ODEforu}) can be expressed in terms of confluent
hypergeometric functions,
\begin{equation}
\textstyle
u(x)={}_1F_1\left(-\frac{b}{4},\frac{1}{2},-x^2\right) +
\beta\,x\,_1F_1\left(\frac{2-b}{4},\frac{3}{2},-x^2\right)
\end{equation}
and is strictly positive for $b>-2$ and 
$|\beta|<2\Gamma(\frac{b}{4}+1)/\Gamma(\frac{b+2}{4})$, cf.\ ref.\
\cite{RJ98a,RJ98b}. 
The corresponding partner potential is given by
\begin{equation}
  V_-(x)=\frac{1}{2}\, x^2 -\frac{b+1}{2}+\frac{u'(x)}{u(x)}\left[2\, x +
  \frac{u'(x)}{u(x)} \right]
\end{equation}
and plots of it for various values of the parameters $b$ and $\beta$
can be found in Figure 1 of \cite{RJ98a} and Figures 1 and 2 of
\cite{RJ98b}. Here we only note that for large $|x|$ this potential becomes
asymptotically that of a harmonic oscillator. For values of $b$ close to the
lower bound $-2$ this potential exhibits two double-wells near the
origin. Whereas for larger values of $b$ these double wells merge to a single
well at the origin. For $\beta=0$ the potential $V_-$ is symmetric about
$x=0$. This symmetry is broken for non-vanishing $\beta$. Noting that SUSY is
unbroken for all allowed values of the parameters, the spectral properties of
the corresponding 
Hamiltonian $H_-$ are easily obtained from (\ref{trafounbroken}):
\begin{equation}
\begin{array}{l}
E_0^-=0\;,\qquad E_{n+1}^-=E_n^+=n+b/2+1 \;, \qquad\displaystyle
\psi_0^-(x)=\frac{\psi_0^-(0)}{u(x)}\,\exp\{-x^2/2\}\;,\\[4mm] 
\displaystyle
\psi_{n+1}^-(x)=
\frac{\exp\{-x^2/2\}}{\left[\sqrt{\pi}\,2^{n+1}n!(n+b/2 + 1)\right]^{1/2}}
\left(H_{n+1}(x)+H_n(x)\,\frac{u'(x)}{u(x)}\right).
\end{array}
\label{result1}
\end{equation}

These results also allow a
complete study of the Fokker-Planck equation for the drift potential 
\begin{equation}
  U(x)=\frac{1}{2}\,x^2+\log u(x)\;.
\end{equation}
In fact, this drift potential is stable and has a stationary distribution
given by $[\psi_0^-(x)]^2$. The decay rates and decay modes are given
explicitly by $E_{n+1}^-$ and $\psi_{n+1}^-$ in
(\ref{result1}). Plots of this drift potential for various values of the
parameters can be found in Figures 1 of \cite{HZ82}. Here we
again briefly mention that for $b$ close to its lower 
limit $U$ has the shape of a bistable (double-well) potential being symmetric
about the origin for $\beta=0$. For larger values of $b$ this drift potential
develops also a stable single-well shape.

To conclude this subsection let us mention that the results presented here are
not new. They have first been derived by Hongler and Zheng
\cite{HZ82} in 1982, which was even before the work of Mielnik
\cite{M84} who was searching for more general factorizations of the harmonic
oscillator Hamiltonian. His results can be viewed as special cases of those
presented here. The advantage of the present approach is that it can be
applied not only to the harmonic oscillator-like systems but to all exactly
solvable ones \cite{RJ98b}. In particular, there exist also shape-invariant
quantum systems which allow for a broken SUSY \cite{Ju96}. That is, one may be
able to design drift potentials which are metastable and can be solved
exactly. To our knowledge, such exactly solvable systems are not available so
far. Except, of course, the rather trivial case of a piecewise linear drift
potential. In the next subsection we are going to present an exactly solvable
metastable drift potential, which is related to the radial harmonic oscillator
Hamiltonian. \\[2mm]

%%%%%%%%%%%%%%%%%%%%%%%%%%%%%%%%%%%%%%%%%%%%%%%%%%%%%%%%%%%%%%%%%%%%%%%%%%%%%%
\noindent{\bf 4.2 \quad A system with broken SUSY}\\[2mm]
In this subsection we will consider the case where the SUSY potential is given
by 
\begin{equation}
 \Phi(x)=x+\frac{\gamma}{x}\;,\qquad \gamma>0\;,
\label{5.1}
\end{equation}
where the condition put on the parameter $\gamma$ leads to a SUSY potential
$\Phi$ which characterizes a shape-invariant SUSY pair of radial
harmonic oscillator-like Hamiltonians with broken SUSY. 
Due to the condition (\ref{ODEforu}) even the SUSY potential (\ref{W}) with
above $\Phi$ gives rise to the radial harmonic oscillator potential 
\begin{equation}
  \label{5.2}
 V_+(x)=\frac{x^2}{2}+\frac{\gamma(\gamma-1)}{2x^2}+\gamma+\frac{b+1}{2}\;,
\end{equation}
which is exactly solvable and leads to the following spectral properties
of $H_+$:
\begin{equation}
E_n^+=2n+2\gamma+1+\frac{b}{2}\;,\quad
\psi_n^+(x)=
\left[\frac{2\,n!}{\Gamma(n+\gamma+1/2)}\right]^{1/2}x^{\gamma}\,
\e^{-x^2/2}\,L_n^{(\gamma-1/2)}(x^2)\;.
\label{dataforH+broken}
\end{equation}
Here $L_n^{(\nu)}$ denotes an associated Laguerre polynomial of degree $n$ and 
index $\nu$. As in the previous case the general solution of (\ref{ODEforu})
can be expressed in terms of confluent hypergeometric functions. Here,
however, we are only interested in cases with broken SUSY and with this
constraint the most general solution reads
\begin{equation}
\textstyle
  u(x)={}_1F_1(-\frac{b}{4},\gamma+\frac{1}{2},-x^2)
      =\e^{-x^2}{}_1F_1(\gamma +\frac{b+2}{4},\gamma+\frac{1}{2},x^2)\;.
\end{equation}
This solution will be strictly positive if $b>-4\gamma-2$. The corresponding
SUSY partner potential reads
\begin{equation}
  V_-(x)=\frac{x^2}{2}+\frac{\gamma(\gamma +1)}{2x^2}+\gamma-\frac{b+1}{2}+
\frac{u'(x)}{u(x)}\left(2x+\frac{2\gamma}{x}+\frac{u'(x)}{u(x)}\right)
\end{equation}
and is plotted, for example, as Figure 5 in ref.\ \cite{RJ98b}. The
eigenvalues of the Schr\"odinger Hamiltonian $H_-$ for this potential are
identically to those of $H_+$ given in (\ref{dataforH+broken}). The
eigenfunctions can also be obtained from those in (\ref{dataforH+broken}) via
the SUSY transformation (\ref{trafobroken}) and read
\begin{equation}
  \psi_n^-(x)=
\left[\frac{2\,n!}{(n+\gamma+\frac{1}{2}+\frac{b}{4})
\Gamma(n+\gamma+1/2)}\right]^{1/2}
x^{\gamma+1}\,\e^{-x^2/2}
\left(L_n^{(\gamma+1/2)}(x^2)+\frac{u'(x)}{2\,x\,u(x)}\right).
\end{equation}

Again, as in the previous case we can construct a drift potential with decay
rates and modes given by the eigenvalues and eigenfunctions of $H_-$. This
drift potential is explicitly given by
\begin{equation}
\textstyle
  U(x)=\frac{1}{2}\,x^2+\gamma\log x +\log u(x)=-\frac{1}{2}\,x^2+\gamma\log x
  + \log\left[{}_1F_1(\gamma +\frac{b+2}{4},\gamma+\frac{1}{2},x^2)\right]
\label{U2}
\end{equation}
and because of broken SUSY does not have a stationary distribution. To be
more explicit, for small $x>0$ it has a logarithmic ``hole'' at the origin,
i.e.\ $U(x)\approx -\gamma|\log x|$ for $x\ll 1$, whereas for large
$x\to\infty$ it becomes asymptotically that of a harmonic oscillator,
i.e.\ $U(x)\approx x^2/2$ for $x\gg 1$. In addition to that, for values of $b$
close to its lower limit $-4\gamma-2$ this drift potential exhibits a local
minimum. In other words, it is metastable. In fact, the smallest decay rate is
given by $E_0^-=2\gamma +1+b/2\searrow 0$ for $b\searrow -4\gamma-2$.
 For larger values of $b$ this local
minimum disappears and the drift potential (\ref{U2}) becomes unstable. A
typical plot of this potential is given in Figure 1 where we have shown
(\ref{U2}) for $\gamma=1$ and $b\in(-6,-5)$. Note
that in Figure 1 we have plotted $U(x)$ versus $\exp\{-x\}$.
\begin{figure}[bt]
\vspace{93mm}%
\includegraphics{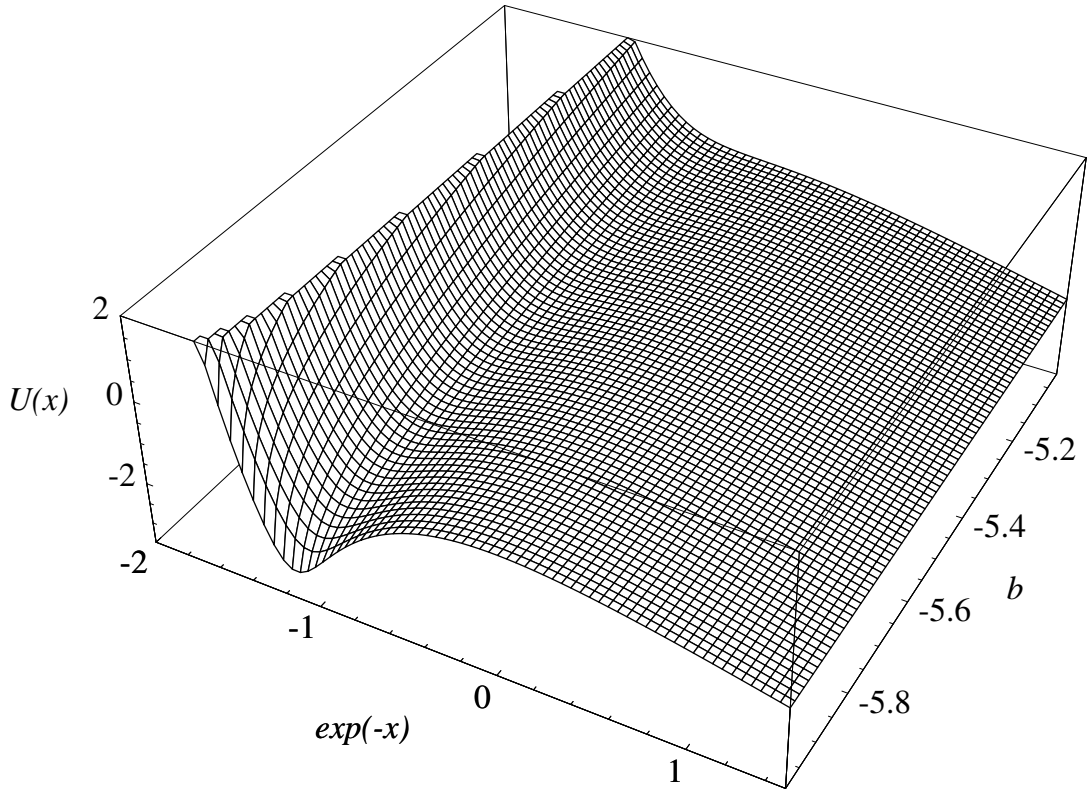}%
\vspace{12mm}%
\caption{A family of drift potentials (\ref{U2}) for $\gamma=1$ showing a
  transition from metastability to instability with increasing $b$. The
  corresponding decay rates and decay modes are known in closed form. Note
  that we have plotted $U(x)$ versus $\exp\{-x\}$.}
\end{figure}%
\\[2mm]

%%%%%%%%%%%%%%%%%%%%%%%%%%%%%%%%%%%%%%%%%%%%%%%%%%%%%%%%%%%%%%%%%%%%%%%%%%%%%%
\noindent{\large\bf 5 \quad Final remarks}\\[2mm]
In this paper we have extended a recent approach to the construction of drift
potentials for which the 
associated Fokker-Planck equation can be solved exactly. As starting point we
have chosen a SUSY potential $\Phi$ which generates a pair of shape-invariant
quantum mechanical potentials leading to exactly solvable quantum
Hamiltonians. This SUSY potential can be perturbed (by adding $u'/u$) in such
a way that one of these Hamiltonians, $H_+$, remains in the class of
shape-invariant exactly solvable Hamilton operators. The partner Hamiltonian
$H_-$, however, is a new one and due to SUSY its spectral properties can be
obtained from $H_+$ in a straightforward way. Using the close relations
between SUSY quantum mechanics and the Fokker-Planck equation we have used
these results to find also new drift potentials with exactly known decay rates
and modes. In this paper we have constructed stable as well as unstable drift
potentials associated with unbroken and broken SUSY, respectively. The examples
for unbroken SUSY are actually not new and have already been studied by
Hongler and Zheng \cite{HZ82}. However, the present results for broken
SUSY leading, in particular, to metastable drift potentials are new. To our
knowledge, these are the first (on the positive half line) analytical
metastable drift potentials for which the decay rates and modes can be given in
closed analytical form. Besides some practical applications these metastable
potentials can also serve as a testing ground for approximation
methods. Recall that for metastable drift potentials the fluctuation operator
for the classical ``bounce'' solution has a negative eigenvalue and,
therefore, 
leads to serious singularities within a saddle-point approximation
\cite{Co77}. The standard treatment in such unstable situations goes back to
Langer \cite{La67} and is based on some analytical continuation techniques. 
These formal treatments can now be applied to and (for the first time) tested
with the examples presented in Section 4.2.\\[2mm]

%%%%%%%%%%%%%%%%%%%%%%%%%%%%%%%%%%%%%%%%%%%%%%%%%%%%%%%%%%%%%%%%%%%%%%%%%%%%%%
\noindent{\large\bf Acknowledgements}\\[2mm]
First of all I would like to thank George Pogosyan for his kind invitation to 
participate in this meeting, which has been very inspiring. Furthermore, I
would like to thank Arousyak and Khajak Karayan for their help, kind
hospitality and many enjoyable conversations. I have enjoyed some
clarifying discussions with Peter H\"anggi and Ulrich Weiss. My thanks go also
to Peter M\"uller for providing me with hard copies of some relevant
papers and Hajo Leschke for valuable comments. Financial 
support by the Heisenberg-Landau program and the Deutsche
Forschungsgemeinschaft is also gratefully acknowledged.
%%%%%%%%%%%%%%%%%%%%%%%%%%%%%%%%%%%%%%%%%%%%%%%%%%%%%%%%%%%%%%%%%%%%%%%%%%%%%%

\end{document}